# Metamagnetism, quantum criticality, hidden order and crystal electric fields in URu$_2$Si$_2$

N. Harrison[a],[*] K. H. Kim[a], M. Jaime[a], J. A. Mydosh[b]

[a]*National High Magnetic Field Laboratory, LANL, MS-E536, Los Alamos, New Mexico 87545*

[b]*Max Planck Institute for Chemical Physics of Solids, Nöthnitzer Strasse 40, D-01187 Dresden, Germany*

**Abstract**

This paper presents a brief synopsis of magnetization, electrical transport, specific heat measurements as well as other recent work on URu$_2$Si$_2$, together with some topical discussions of the groundstate properties in relation to metamagnetism, quantum criticality and crystal electric fields. © 2001 Elsevier Science. All rights reserved

Among heavy fermion systems, the groundstates of URu$_2$Si$_2$ are particularly complex. Upon cooling, prior to becoming superconducting at ~ 1 K, this material undergoes a second order phase transition into an ordered phase at $T_o$ = 17.5 K [1], the nature of which remains controversial [1-14]. Owing to the absence of a microscopic description or origin, the term "hidden order" has been coined. Some of its physical properties have recently been proposed to be consistent with unconventional forms of density-waves involving a *d*-wave pairing manifold [6,9].

Early experimental data on this system were interpreted in terms of crystal electric field effects involving at least two lowest energy $J_z$ = 0, $J$ = 4 singlet states [3]. The primary evidence that supported such an interpretation was found by means inelastic neutron scattering [4]: these experiments revealed a gap $\Delta$ associated with magnetic excitations polarized along the tetragonal *c*-axis, in which the levels appeared not to be broadened in a linear fashion by the magnetic field as would be expected in the case of doublets. A scheme of levels was devised that could explain the existence of these excitations, as well as the maximum in the susceptibility at ~ 50 K [3], the possibly of a weak Schottky anomaly in the specific heat at ~ 30 K (albeit that model-specific fitting was never actually performed) [5] and metamagnetism resulting from crossings of the crystal field levels at high magnetic fields [3,5]. The existence of small splittings (< 10 meV) between the lowest lying crystal field levels therefore appeared to make the underlying physics of URu$_2$Si$_2$ different from most other heavy fermion systems. The same set of levels also provided the basis set for postulating unconventional forms of electric antiferroquadrupolar order [5].

In spite of these early advances in understanding URu$_2$Si$_2$, the crystal symmetry-breaking structural transformation that should accompany a phase transition into an antiferroquadrupolar ordered phase was never found. Furthermore, the narrowing and shifting of the proposed Schottky-anomaly-like-feature to lower temperatures with magnetic field, that should occur for the proposed level scheme on the approach to metamagnetism, has not been observed [10]. More recently, inelastic neutron scattering studies extended to magnetic fields of order ~ 17 T revealed the characteristic gap $\Delta$ to increase with magnetic field [11] in contrast to the decrease anticipated by crystal field models [5]. This latter finding has two important implications for URu$_2$Si$_2$: firstly, the original proposed scheme of crystal field levels is incorrect, and, secondly, the metamagnetism does not originate from the crossings of such levels in a magnetic field. Rather, the metamagnetism at 7 and 8 K shown in Fig. 1 (from Reference [12]) appears to be more reminiscent of that seen in itinerant electron metamagnets such as $CeRu_2Si_2$ [15], $UPt_3$ [16] and $Sr_3Ru_2O_7$ [17]. Notably, the size of the saturated moment in $URu_2Si_2$ is half the value (~ 3 $\mu_B$) anticipated by crystal field models [3], yet similar to that (~ 1.5 $\mu_B$) observed in $CeRu_2Si_2$ [15]. There are therefore good reasons now to believe that the underlying heavy fermion state of $URu_2Si_2$, from which the hidden order condenses, is more similar to other heavy fermion systems than was previously thought. Clearly, alternative models for the origin of the low energy magnetic excitations found in inelastic neutron scattering studies need to be considered. The recent finding that the hidden order groundstate of $URu_2Si_2$ competes with antiferromagnetism may provide such an alternative model [7]. Numerous experiments now show that the antiferromagnetic phase of $URu_2Si_2$ appears via a first order transition with phase separation under pressure [13], and possibly also as an impurity phase (with ~ 1 % volume fraction) at ambient pressure [7], where the bulk antiferromagnetic phase of $URu_2Si_2$ is separated in energy from the bulk hidden order phase by a gap very comparable in value to $\Delta$. Inelastic neutron scattering studies [4] could then imply that the spectrum of magnon excitations normally present within the antiferromagnetic phase becomes gapped within the hidden order phase. While antiferromagnetic correlations must still be present within the hidden order phase, the hidden order parameter appears to exclude (or, at least, be incompatible with) bulk long range antiferromagnetic order.

---
[*]Corresponding author. Tel.: +1-505-6653200; fax: +1-505-6654311; e-mail: nharrison@lanl.gov

In spite of the fact that the hidden order phase seems to be one in which the local moments are fully quenched, the parent antiferromagnetic phase of $URu_2Si_2$ should be compared to other heavy fermion antiferromagnets. $UPd_2Al_3$ could be a suitable paradigm [18]. The incipient antiferromagnetism combined with superconductivity places $URu_2Si_2$ slightly to the left-side of the zero magnetic field quantum critical point in the generic heavy fermion phase diagram sketched in Fig. 2. $UPt_3$ [19] and, in particular, $CeRu_2Si_2$ [20] are considered to be close to quantum criticality at ambient pressure. In all cases, such close proximity to a quantum critical point makes these materials susceptible to forming new ordered phases as a means of lowering energy. Perhaps this is achieved in $URu_2Si_2$ by the formation of both the hidden order phase and superconductivity [1]. Since the hidden order phase in $URu_2Si_2$ competes with antiferromagnetism, it probably occupies a very narrow region in $x$ or $p$ just inside the antiferromagnetic phase in Fig 2, but that extends to rather high temperatures and magnetic fields.

The application of a magnetic field to $URu_2Si_2$ has the potential to produce various types of quantum criticality, either from the suppression of the parent antiferromagnetic phase, the hidden order phase or by the induction of itinerant electron metamagnetism. New physics should therefore be expected and this is indeed what can be seen in Fig. 3a. The existence of many competing interactions in $URu_2Si_2$ leads to a complex phase diagram as a function of the applied magnetic field that is unique. The version of the phase diagram shown in Fig. 3a is compiled from resistivity $\rho$ measurements made both as a function temperature and field [14]. Phase transitions that are present in the magnetization [12], specific heat [10] or ultrasound velocity [21] lead to extrema in the first derivatives $d\rho/dT$ and $d\rho/dH$. The formation of multiple phases can be shown to be connected with the existence of a field-induced quantum critical point at $37 \pm 1$ T. This becomes apparent in Fig. 3b upon eliminating the regions in $H$ versus $T$ space occupied by ordered phases. At magnetic fields higher than ~ 39 T, low temperature fits of the function $\rho = \rho_0 + aT^n$ to resistivity $T$-sweeps yield Fermi liquid behaviour for which $n \sim 2$ only below a characteristic temperature $T^*$. The Fermi liquid parameter $a$ appears to diverge on the approach to $\mu_0H_0 \sim 37$ T, evidenced by the fact that a fit of $a^{-1/2}$ (which is proportional to $1/\gamma$ where $\gamma$ is the Sommerfeld coefficient) intercepts the $T = 0$ axis in a $|H-H_0|^\alpha$ power law-like fashion where $\alpha \sim 0.7$: (actual values of $a^{-1/2}$ are plotted as × symbols). At temperatures above $T^*$ (open circles), the resistivity crosses over to a non-Fermi liquid behaviour characterized by $n \leq 1$: ($T^*$ can be obtained explicitly by observing the maximum in $d\rho/dT$ plotted versus $T$). A power law fit to $T^*$ once again intercepts the field axis at $\mu_0H_0 \sim 37$ T, with $\alpha \sim 0.6$. At fields lower than 37 T and temperatures above ~ 7 K, the magnetoresistance undergoes a broad maximum: a power law fit of its $T$ versus $H$ loci (filled circles) also intercepts the field axis at $\mu_0H_0 \sim 37$ T where, this time, $\alpha \sim 0.5$.

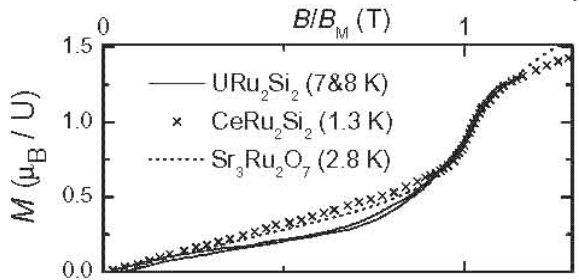

Fig.1. A comparison of itinerant electron metamagnetism in three compounds [15,17] (including $URu_2Si_2$) where $B_M = \mu_0 H_M$ denotes the field at which the transition or crossover occurs. Only in the case of $Sr_3Ru_2O_7$ is the value of the moment renormalized.

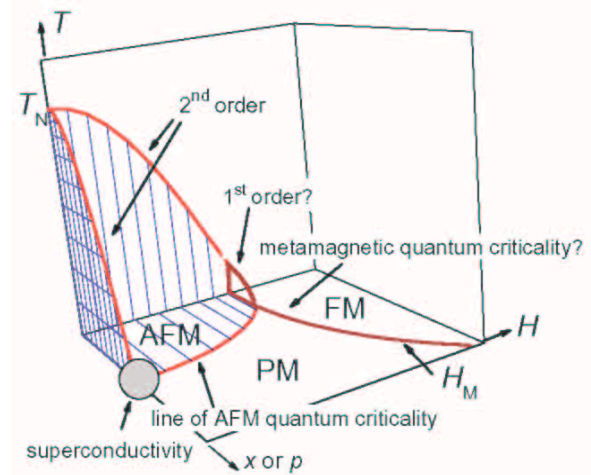

Fig.2. Possible generic phase diagram of heavy fermion compounds, generalized to include magnetic field. AFM, FM and PM refer to antiferromagnetic, ferromagnetic and paramagnetic respectively. The hidden order phase, which occurs only in $URu_2Si_2$, is not shown. $x$ and $p$ refer to doping and/or pressure which may be offset with respect to zero, depending on the material. At ambient pressure, $UPt_3$ and $CeRu_2Si_2$ are close to exhibiting a quantum critical point at $B = 0$, possibly occupying the grey circle region. Because of its incipient antiferromagnetism, ambient pressure $URu_2Si_2$ probably occurs at slightly lower $x$ or $p$ where it is overcome by superconductivity.

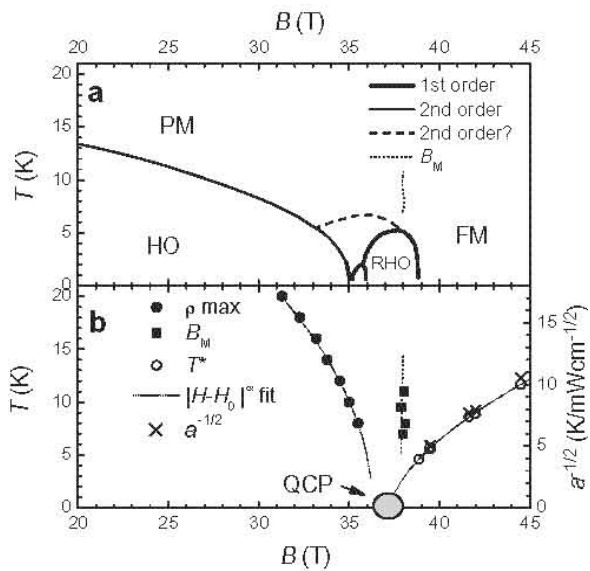

Fig. 3. (a) The phase diagram of $URu_2Si_2$ as a function of magnetic field and temperature extracted from Reference [14] in which actual data points are omitted for clarity. HO and RHO refer to hidden order and reentrant hidden order [12] respectively. The two other newly discovered ordered phases are not labeled. (b) An assemblage of physical properties of $URu_2Si_2$ outside the ordered phases, together with power law fits as described in the text. Note that $T^*$ overlays $a^{-1/2} \propto \gamma^{-1}$ on selecting appropriate vertical axis scaling, suggesting that $T^*$ is like an effective Fermi bandwidth. All data points extrapolate to a metamagnetic quantum critical point (QCP) at $37 \pm 1$ T.

While broad magnetoresistivity maxima may not be a generic feature of all field-induced quantum critical points, they are recognized as a feature specific to metamagnetic quantum critical end points. Such points are accompanied by a divergence in $a$, vanishing of $T^*$, divergence in $\gamma$ and divergence in the susceptibility $\chi$. An accurate determination of $\gamma$ at high magnetic fields has yet to be made. Measurements of $\chi$ in pulsed magnetic fields do, however, exhibit a divergent-like $T$-dependent behaviour at temperatures above ~ 6 K in Figs. 4a and 4b [12]: the low temperature portion is cut off by the formation of the reentrant hidden order phase. The magnetization measurements featured in this paper reveal that the metamagnetic crossover occurs at $\mu_0H_M \sim 37.8$ T, so that it approximately coincides with the extrapolated magnetoresistance maximum, $T^*$ and $a^{-1/2}$ only at $T = 0$. The term `metamagnetic crossover' is used at finite temperatures because a true first order phase transition is thought only to exist at $T = 0$. All of the above evidence for quantum criticality appears to be associated with metamagnetism, rather than with the suppression of antiferromagnetism or the hidden order phase (found to terminate at $H_c \sim 35$ T). In fact, the increase in the excitation gap $\Delta$ with field revealed by inelastic neutron scattering studies [11] implies that antiferromagnetism becomes more energetically unfavourable, making the observation of antiferromagnetic quantum criticality unlikely.

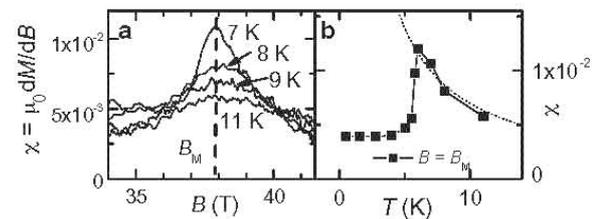

Fig. 4. (a) The susceptibility on passing through the metamagnetic crossover at elevated temperatures in $URu_2Si_2$. (b) The temperature-dependence of the $\chi$. The dotted line is a plot of the function $k/T$ (where $k$ is a constant) showing that $\chi$ appears to be diverging prior to the formation of the RHO phase.

Itinerant electron metamagnetic quantum criticality is different from the better known antiferromagnetic quantum criticality in that no symmetry breaking order parameter is involved [22]. Since the relevant fluctuations in the system involve the magnetization, Millis *et al.* [22] proposed treating the expectation value of the magnetization as a quantity equivalent to the expectation value of an order parameter. The concept of a line of first order phase transitions terminating at a quantum critical end point at a critical field $B_M$ (and field orientation) seems to work well for $Sr_3Ru_2O_7$, which is predisposed to be ferromagnetic [17,23]. However, the origin of the first order quantum critical end point in heavy fermion systems remains something of a mystery. In these systems, metamagnetism can be effectively explained by a renormalized band-like picture, in which *f*-electron states (shifted by a large self energy) hybridize with conduction electrons [24]. The dispersionless properties of the *f*-electrons naturally produce a single step-like feature in the magnetization at a critical field that depends on the Kondo temperature scale. However, the absence of an actual level crossing at $B_M$ is uncharacteristic of first order phase transitions. One interesting possibility is that another parameter in the system, such as the Kondo-liquid condensation temperature [25] or hybridization parameter $V$ [24], plays the role of an order parameter. Differences in the values of either of these quantities above and below $B_M$ have the potential to yield two simultaneous values for the free energy at fields around $B_M$, giving rise to a level crossing and hence a first order phase transition at finite or zero temperature.

While key questions regarding metamagnetism still need to be answered, it appears to be the case that the multiple phases shown in Fig. 3a are a consequence of the interplay between the hidden order parameter and metamagnetic quantum criticality. This is especially evident in the susceptibility: Fig. 5 shows a contour plot of $\chi$ with symbols delineating maxima that can be interpreted as transitions or crossovers. A single metamagneticcrossover at high temperatures splits into two transitions below $\sim 6$ K, giving rise to the domed "reentrant" hidden order (RHO) phase [12]. Transport studies performed in static magnetic fields provided by the hybrid magnet in Tallahassee [14] reveal the reentrant phase (phase III) to be first order at all temperatures. This is evidenced by hysteresis in the position of the transitions that depends on the direction in which $T$ or $H$ is swept. All transitions, including the upper critical field of the hidden order phase $H_c$ become strongly hysteretic below $\sim 3$ K. Fig. 6 shows the hysteresis in low temperature resistivity measurements observed upon sweeping the field.

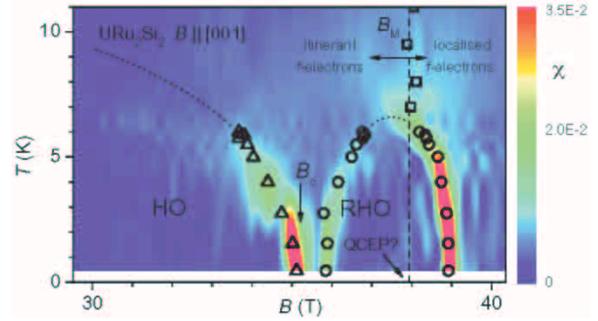

Fig. 5. Intensity plot of the susceptibility $\chi$ of $URu_2Si_2$ measured in pulsed magnetic fields as a function of temperature. Symbols indicate maxima that can be either first order phase transitions at low temperatures or a metamagnetic crossover at high temperatures.

The complexity of the phase diagram and the magnetic excitation spectrum obtained in inelastic neutron scattering experiments ought to provide important clues as to the nature of the hidden order parameter. For example, the appearance of a magnon excitation gap at (1,0,0) in the antiferromagnetic Brillouin zone, that is present only in the hidden order phase but that is commensurate with the Bragg peaks of the antiferromagnetic phase, is consistent with a scenario in which the hidden order parameter excludes antiferromagnetic order. This might appear to support suggestions that the hidden order phase involves antiferroquadrupolar ordering of $\Gamma_5$ doublets, since antiferromagnetic and antiferromagnetic order parameters do not generally coexist [7]. However, neither a change in crystalline structure nor the emergence of magnetic field-induced dipolar structure characteristic of antiferroquadrupolar ordered states formed from doublets [26], has been observed. An alternative explanation for the inability of the hidden order and antiferromagnetic phases to spatially coexist could be that the former involves predominantly itinerant *f*-electrons while the latter involves localized *f*-electrons. This is the kind of thing that would be expected for the novel *d*-wave density-wave models of the hidden order parameter [6,9].

The manner in which the primary hidden order phase evolves in a magnetic field should also provide important clues. Possible candidates for its destruction, include (i) a spin-flip-like process, like those in conventional antiferromagnetically ordered systems, (ii) the destruction of the order parameter induced by abrupt Fermi surface topology changes associated with metamagnetism as the *f*-electrons transform from itinerant to localized ferromagnetic behaviour [24], or (iii), Pauli-limiting of a spin-singlet order parameter. The apparent absence of local moments within the hidden order phase makes it difficult to imagine how (i) can be relevant in this system. Meanwhile, substitution studies of Rh in place of Ru reveal that the hidden order upper critical field $H_c$ and the metamagnetic crossover $H_M$ separate widely in field [27]. This implies that the two transitions are uncorrelated, making (ii) unlikely. The value of $H_c$ is, however, consistent with the Pauli limit of a spin-singlet order parameter upon making a BCS estimate of the gap from $T_o$ [12], suggesting that (iii) may be conceivable. The *d*-wave orbital antiferromagnetic order parameter (involving bond currents between planar U atoms) is proposed to be spin singlet, while the *d*-wave spin-density wave order parameter should normally be spin triplet. The manner in which the HO phase is suppressed by a magnetic field (terminating at a first order phase transition) appears to support the former. However, the incommensurate neutron diffraction Bragg reflection peaks predicted by bond current ordering models have not yet been observed [9,28]. Such experiments may also be resolution limited, since the peaks are predicted to be 50 times lower in intensity than the weak (1,0,0) tiny moment peaks ascribed to the antiferromagnetic impurity phase.

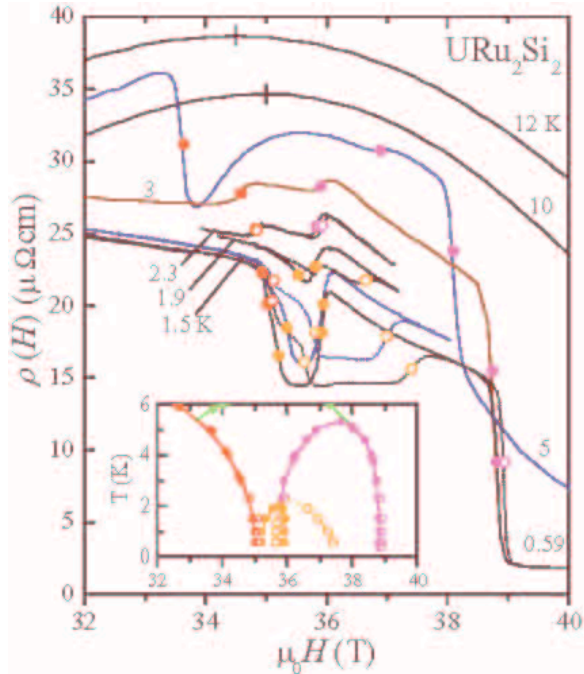

Fig. 6. A portion of the measured resistivity of $URu_2Si_2$ at selected temperatures and intervals in magnetic field for rising (open symbols) and falling (filled symbols) magnetic fields. The '+' symbols indicate the position of the maximum in the magnetoresistance. The inset shows hysteresis in the phase transitions, thus obtained, revealing them to be first order at low temperatures.

The formation of the new ordered phase (or reentrant phase) at $B_M$, where the Fermi surface is extremely asymmetric with respect to spin-up and spin-down *f*-electrons, is further consistent with a spin-singlet order parameter [12]. In order to explain the formation of a plateaux-like feature in the magnetization, the order parameter must open up a gap in the spin-up *f*-electron band, leading to the possible coexistence of localized and itinerant *f*-electrons; possibly alternating between consecutive U planes in the body-centered tetragonal lattice. On condition that the order parameter is spin-singlet, the anisotropy of $H_c$ (the upper critical field of the hidden order phase), with an easy axis along *c*, can be explained by the existence of Ising spin-like quasiparticles similar to those in $CeRu_2Si_2$. de Haas-van Alphen measurements on $URu_2Si_2$ [29] show that significant Zeeman splitting of the spin-up and spin-down itinerant electrons occurs only when the direction of the magnetic field is rotated out of the tetragonal plane.

Conventional spin-density wave groundstates are not usually destroyed by magnetic fields, because they are spin triplet: if anything, magnetic fields tend to improve nesting conditions, leading to a field-induced enhancement of the ordering temperature. An important question that remains to be answered is whether this continues to be true for *d*-wave spin-density wave systems: a theory specific to *d*-wave spin-density wave systems has not yet been developed.

In conclusion, the evidence today supports a picture in which $URu_2Si_2$ starts out as a regular U-based heavy fermion system, with antiferromagnetic correlations. At ambient pressure, antiferromagnetic order is thwarted by an abrupt second order phase transition at ~ 17.5 K into a new type of ordered phase (with broken time-reversal symmetry) that may be unique to $URu_2Si_2$. Direct evidence for translational symmetry breaking, that is required to establish one of the density wave scenarios, has yet to be found. It is essential that the present theoretical models be extended to include both strong magnetic fields and the changes in Fermi surface topology associated with metamagnetism. It is expected that much will be learnt in this system by performing detailed high magnetic field studies under the application of pressure and Rh-doping.

Finally, in Fig. 7, it is instructive to compare the $H$ versus $T$ phase diagram of $URu_2Si_2$ with the generic $x$ (doping) versus $T$ phase diagram of the cuprate superconductors [30]. It is regularly assumed that the pseudogap regime of the cuprates is a true thermodynamic phase characterised by an order parameter $\Delta_p$ that gives rise to quantum criticality when $\Delta_p \rightarrow 0$ at optimal doping $x_{opt}$. An alternative possibility is that the pseudogap behaviour does not correspond to a thermodynamic phase at all, but is the product of quantum fluctuations of a quantum critical end point concealed within the superconducting dome. The existence of strong Coulomb interactions in the cuprates makes them susceptible to electronic structural transitions analogous to the α–γ transition in Ce. Such first order transitions have the potential to become quantum critical in a similar manner to itinerant electron metamagnetism if they terminate at zero rather than finite temperature. Rather than being caused by an ordered phase, pseudogap behaviour could be the consequence of a finite temperature crossover in transport behaviour as is the case for the magnetoresistivity maximum in $URu_2Si_2$ at T > 6 K.

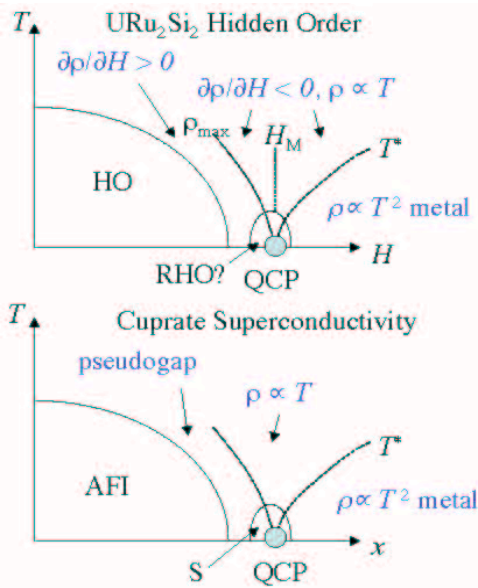

Fig. 7. Cartoon comparing simplified $URu_2Si_2$ and cuprate superconductor phase diagrams exhibiting quantum criticality. AFI and S refer to antiferromagnetic insulator and superconductor respectively.